\documentclass[final]{cimentomod}

\usepackage{amsmath}
\usepackage{amssymb}
\usepackage{graphicx}
\usepackage{epsfig}
\usepackage{pstricks,pst-plot,pst-fill,pst-text}

\newcommand{\ordEW}{\mathcal{O}(\alpha_{\scriptscriptstyle EM}^6)}
\newcommand{\ordQCD}{\mathcal{O}(\alpha_{\scriptscriptstyle EM}^4
  \alpha_{\scriptscriptstyle S}^2)}
\newcommand{\ordQCDtwo}{\mathcal{O}(\alpha_{\scriptscriptstyle EM}^2
  \alpha_{\scriptscriptstyle S}^4)}

\title{Probing EWSB through vector boson scattering at the LHC
\thanks{The original publication is available at Nuovo Cimento Vol.32 C, N. 3-4.}}
\author{D.B. Franzosi\from{ins:to}}
\instlist{\inst{ins:to}INFN and Dipartimento di Fisica Teorica, Universit\`{a} di Torino}

\PACSes{\PACSit{24.10.Lx}{Monte Carlo simulations}
\PACSit{11.15.Ex}{Spontaneous breaking of gauge symmetries}
\PACSit{12.60.-i}{Models beyond the standard model}
}

\begin{document}

\maketitle


\begin{abstract}
We estimate the power of the LHC in probing effects of strongly-interacting symmetry breaking 
sector through vector boson scattering in a complete partonic analysis. 
\end{abstract}

\section{Introduction}

The high energy behaviour of Longitudinal Vector Boson Scattering (LVBS),
$V_LV_L\rightarrow V_LV_L$, reflects the strength of the interactions in the
symmetry breaking (SB) sector.  The SM with a light Higgs predicts a big
suppression of longitudinal boson scattering and a weakly coupled symmetry
breaking sector, while without the Higgs LVBS grows with the scattering
energy $s$ and must be cut off by new strong dynamics at higher
scales~\cite{intro}.

Heavy resonances from the SB sector are possible at LHC energy scale, however
an equally likely signature is simply an excess of events at high energy in
comparison with the SM predictions. This excess of events can be mimicked by
the SM without the Higgs, whose predictions can be compared with more
specific models. For instance, a promising alternative to the complete
replacement of the Higgs mechanism is to interpret the Higgs as a
pseudo-Goldstone boson of a new strongly-interacting sector~\cite{SILH}. This
class of models predict in general a smaller excess than the \emph{No
  Higgs}~(NOH) case.  On the other hand, models with lighter resonances
normally predict a larger excess of events. Residing on this frontier, the
NOH model represents a benchmark scenario for the study of strong SB
dynamics.

A recent study~\cite{munu} has estimated the power of the LHC to distinguish the \emph{No Higgs} 
scenario from the SM through VBS in a complete partonic analysis of the $\mu\nu$+4 jets channel.
This result, complemented with the others important still unpublished
channels, are presented in the following.

The only way to observe VBS at LHC is by looking at the vector bosons through
their decays into fermions.  Therefore, $V_LV_L\rightarrow V_LV_L$ is
necessarily embedded in a more general gauge invariant set comprehending all
processes with six-parton final states. In order to have a good description
of the high energy behaviour of such processes it is necessary to compute the
complete set of Feynman diagrams taking in account all irreducible
backgrounds and their interferences with signal topologies, because of large
gauge cancellations~\cite{gaugecancellation}.  We have analyzes, from this
six-parton perspective, the following channels: $pp\rightarrow 4j\ell\nu$,
$pp\rightarrow 4j\ell^+\ell^-$ and $pp\rightarrow 2j\ell^+\ell^-\ell'\nu$,
where $j$ stands for any quark or gluon and $\ell,\ell'$ for muons or
electrons.  Using the \texttt{PHANTOM}~\cite{phantom} event generator, a MC
dedicated to processes with six partons in the final state, we have generated
the electroweak part, $\ordEW$, for the light Higgs and the \emph{No Higgs}
scenarios, and the $\ordQCD$ QCD background. For the $\ordQCDtwo$ V+4jets QCD
background of the semi-leptonic channels, we have used
\texttt{MADEVENT}~\cite{madevent}.

\section{Complete Partonic Analysis}

The VBS experimental signature can be basically described by the presence of
tag jets, two very energetic jets in the forward-backward direction, and two
pairs of fermions associated with VB decays in the central region, with high
transverse momenta. We have performed a cut-based selection in order to
simultaneously characterize the signal and suppress the background for each
channel, with the aim to optimize the probability to obtain experimental
results outside the $95\%$ confidence region (PBSM@95\% CL) for the SM,
assuming the \emph{No Higgs} scenario.

For tagging jets we required $\Delta\eta(j_fj_b)>4.8$ and
$M(j_fj_b)>1000$~GeV. For the characterization of leptonic vector bosons we
used $|\eta(\ell^\pm)|<2$ and $p_T(V_{rec})>200$~GeV. In the semi-leptonic
channels, the dominant background is $V + 4$~jets, pushing us to a very tight
characterization of the hadronic vector boson: we required $p_T(j_c)>60$~GeV
and $70$~GeV $<M(j_cj_c)<100$~GeV. For the isolation of both reconstructed
bosons we imposed $\Delta\eta(Vj)>0.6$. In addition, a further specialized
set of cuts was applied to each channel to help in distinguishing the
scenarios. For instance, in the $4j\ell\nu$ channel, top rejection is quite
important. The expected number of events for $L=200$~fb$^{-1}$ for each channel
after all cuts are reported in table~\ref{nevents}.

\begin{table}[h!tb]
\begin{tabular}{|l|r|r|r|r|}
\hline
{\bf Channel} 	   & $\ordEW$(SM) & $\ordEW$(NOH) & $\ordQCD$   & $\ordQCDtwo$  \\ 
\hline
$4j\ell\nu$	   & 128.4     	  & 381.6	  & 92  	& 1956  \\
\hline
$4j\ell^+\ell^-$   & 10.6         & 52.8          & 8.8 	& 220   \\
\hline
$2j\ell^+\ell^-\ell'\nu$ & 2.8    & 9.9  	  & 0.8 	& 0   \\
\hline
\end{tabular}
\caption{Expected number of events for $L=200$~fb$^{-1}$ after all cuts, for
  each perturbative order. The contribution from the QCD background is
  essentially the same for the \emph{No Higgs} and the SM case.}
\label{nevents}
\end{table} 

The probability distribution for each scenario was computed by taking the MC
prediction (table~\ref{nevents}) as the mean value and then estimating the
associated uncertainties.  We have assumed a Poissonian statistical
distribution and modeled the theoretical uncertainty due to parton
distributions, scale dependence and higher order corrections with a $30\%$
flat smearing around the mean value.  For the semi-leptonic channels, the
theoretical error for $V+4$~jets totally overwhelms the possibility of
appreciating differences between the scenarios.  Fortunately, at LHC, it will
be possible to get rid of this uncertainty using side-band techniques.  The
mass distribution of the central jets, $M(j_cj_c)$, is characterized by a
peak in the range $70$--$100$~GeV, due to vector boson decays.  The peak is
surrounded by a slowly varying region populated by jet pairs from the $V+4$~jets
background.  This almost flat distribution can be measured and interpolated
to the peak region, providing an estimate of the $V+4$~jets background which is
free of theoretical errors.

\begin{figure}[h!tb]
\centering
\includegraphics*[width=0.32\textwidth,height=4.0cm]{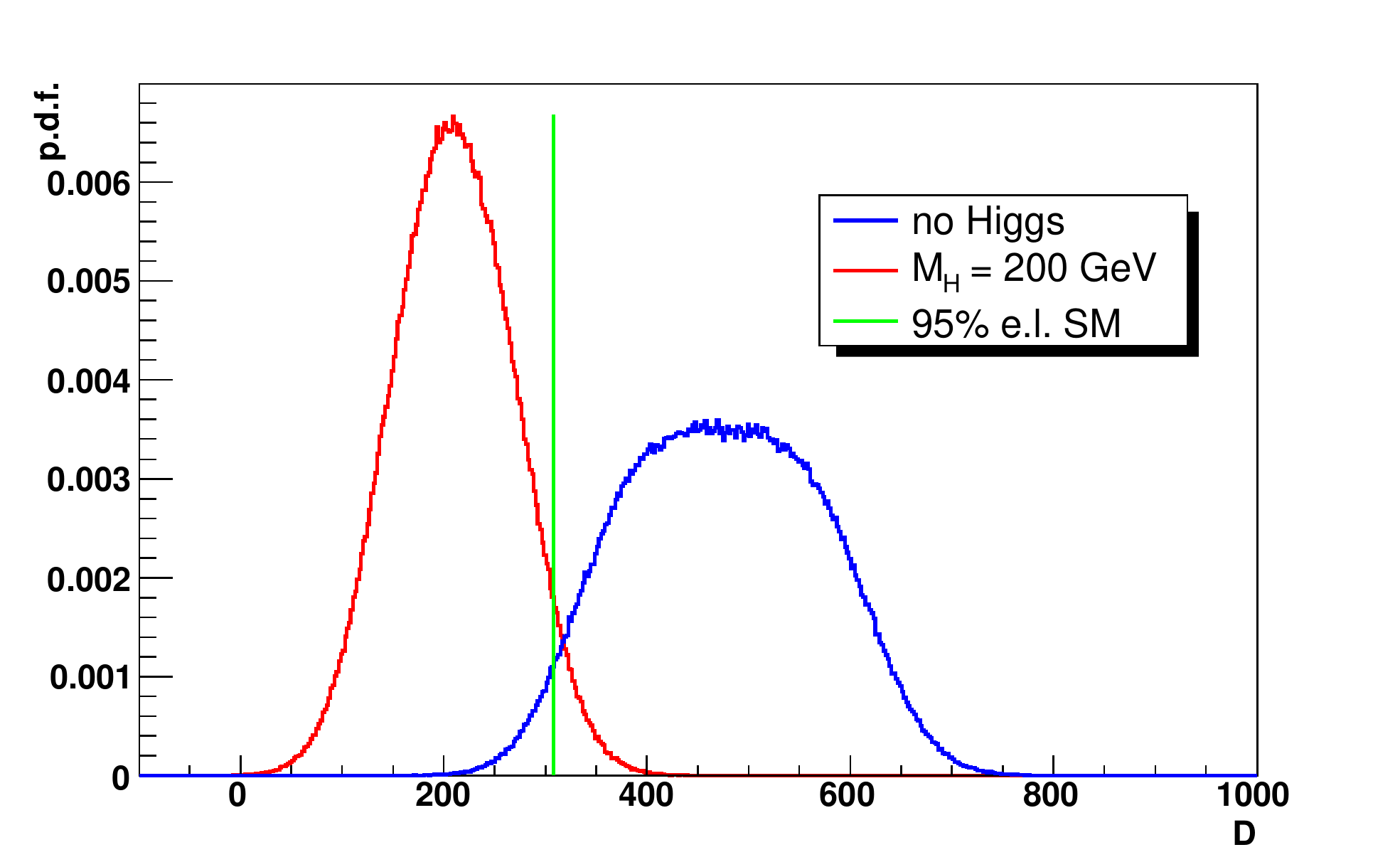}
\includegraphics*[width=0.32\textwidth,height=4.0cm]{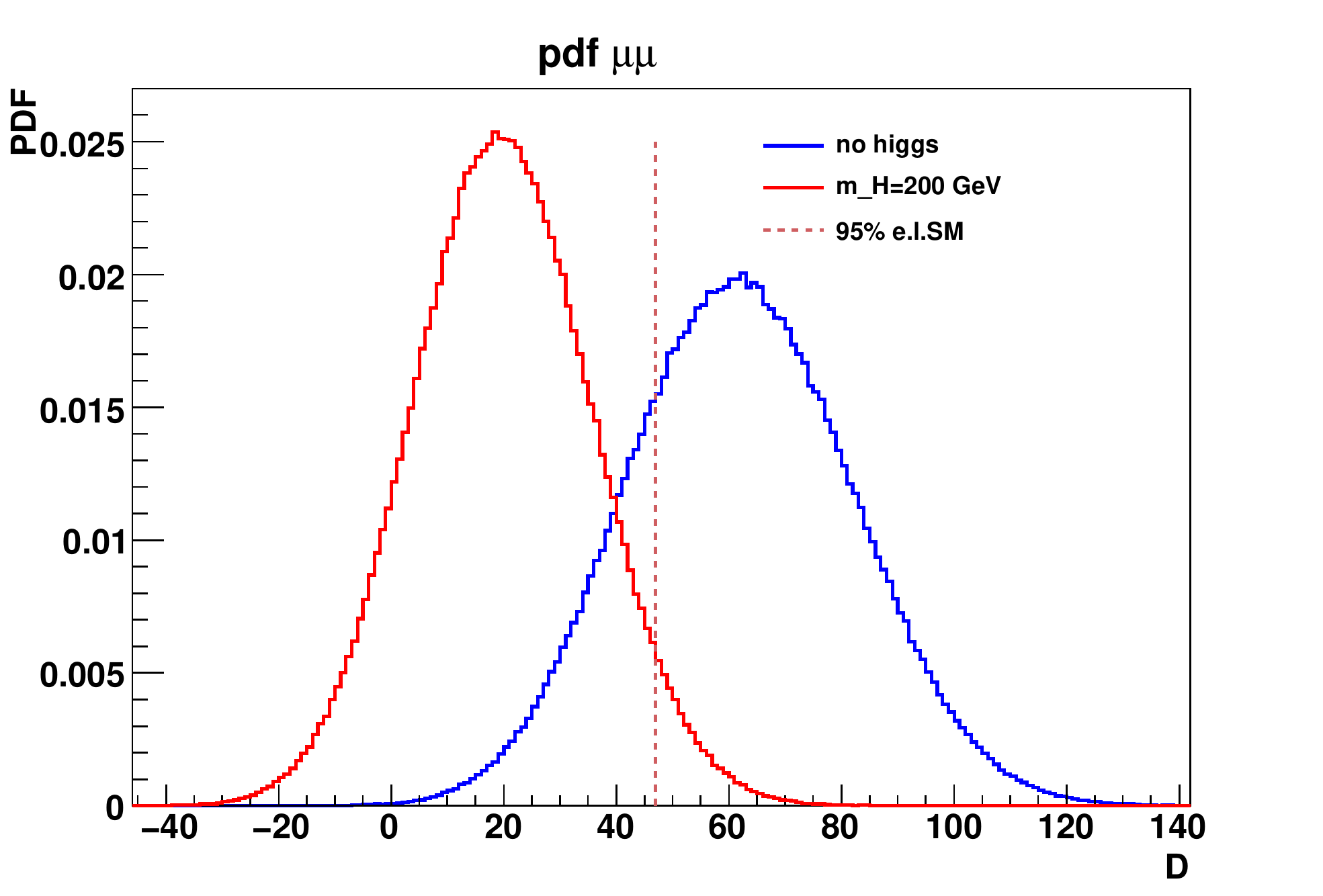}
\includegraphics*[width=0.32\textwidth,height=4.0cm]{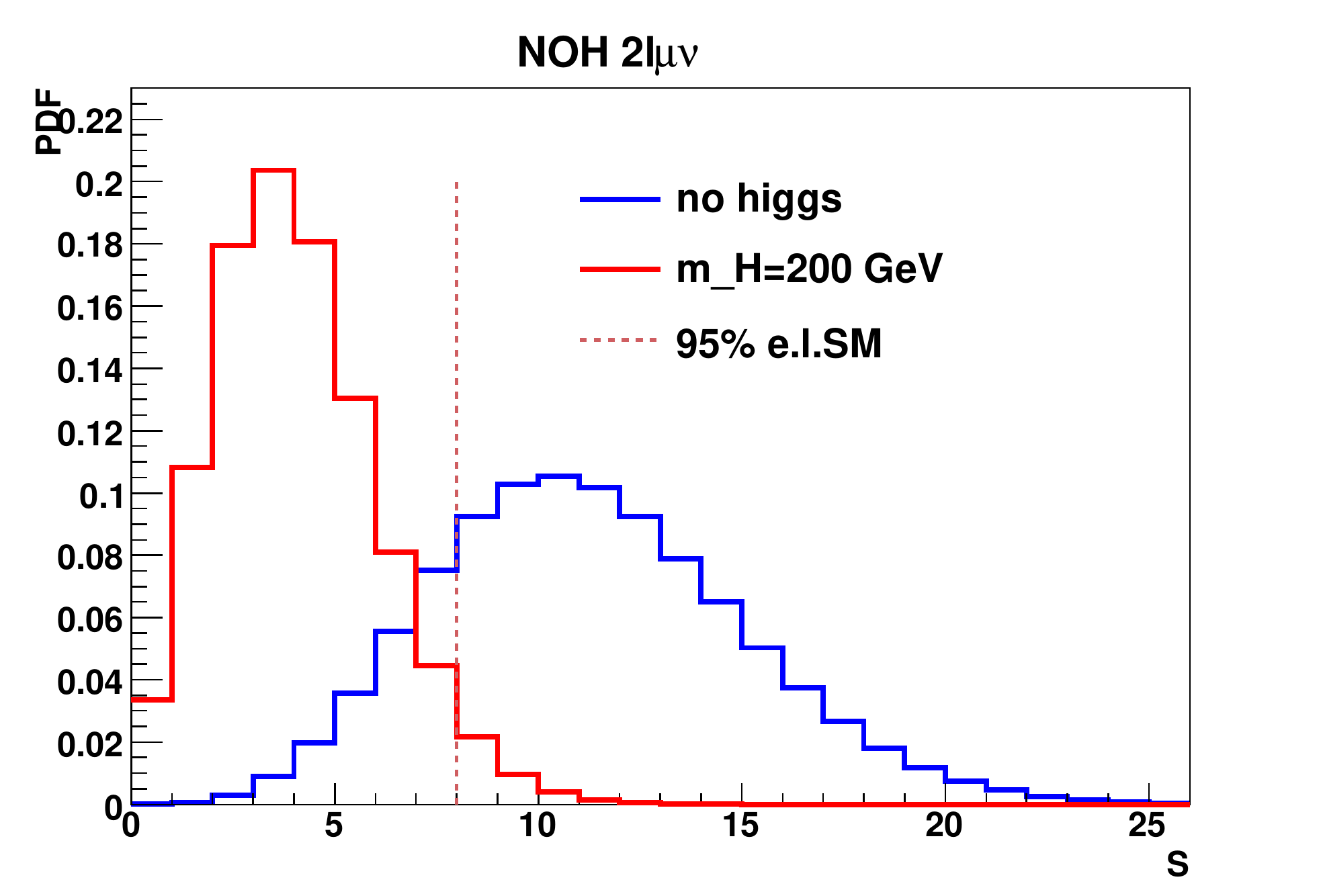}
\caption{Probability distribution of the SM~(red) and \emph{No Higgs}~(blue)
  scenarios. The vertical line is the 95\% exclusion limit of the SM. The
  PBSM@95\% CL is reported in parenthesis. The channels, from left to right,
  are: $4j\ell\nu$~(96.8\%), $4j\ell^+\ell^-$~(77.1\%) and
  $2j\ell^+\ell^-\ell'\nu$~(80.0\%).}
\label{pdf}
\end{figure}

Finally, we defined a discriminator $D$, which estimates the excess of events
above the measured side-bands taking in account the theoretical and
statistical errors of $\ordEW+\ordQCD$ and the statistical errors of
$V+4$~jets background. For the leptonic channel we directly used the number
of events subject to all sources of errors.  Fig.~\ref{pdf} shows the
probability distributions for $L=200$~fb$^{-1}$ of the \emph{No Higgs} and SM
scenarios.  The vertical line represents the 95\% exclusion limit for the SM,
which defines the PBSM@95\% CL for the \emph{No Higgs} scenario.  The
PBSM@95\% CL is reported in the caption of Fig.~\ref{pdf}.

\begin{figure}[b]
\centering
\includegraphics*[width=0.32\textwidth,height=4.0cm]{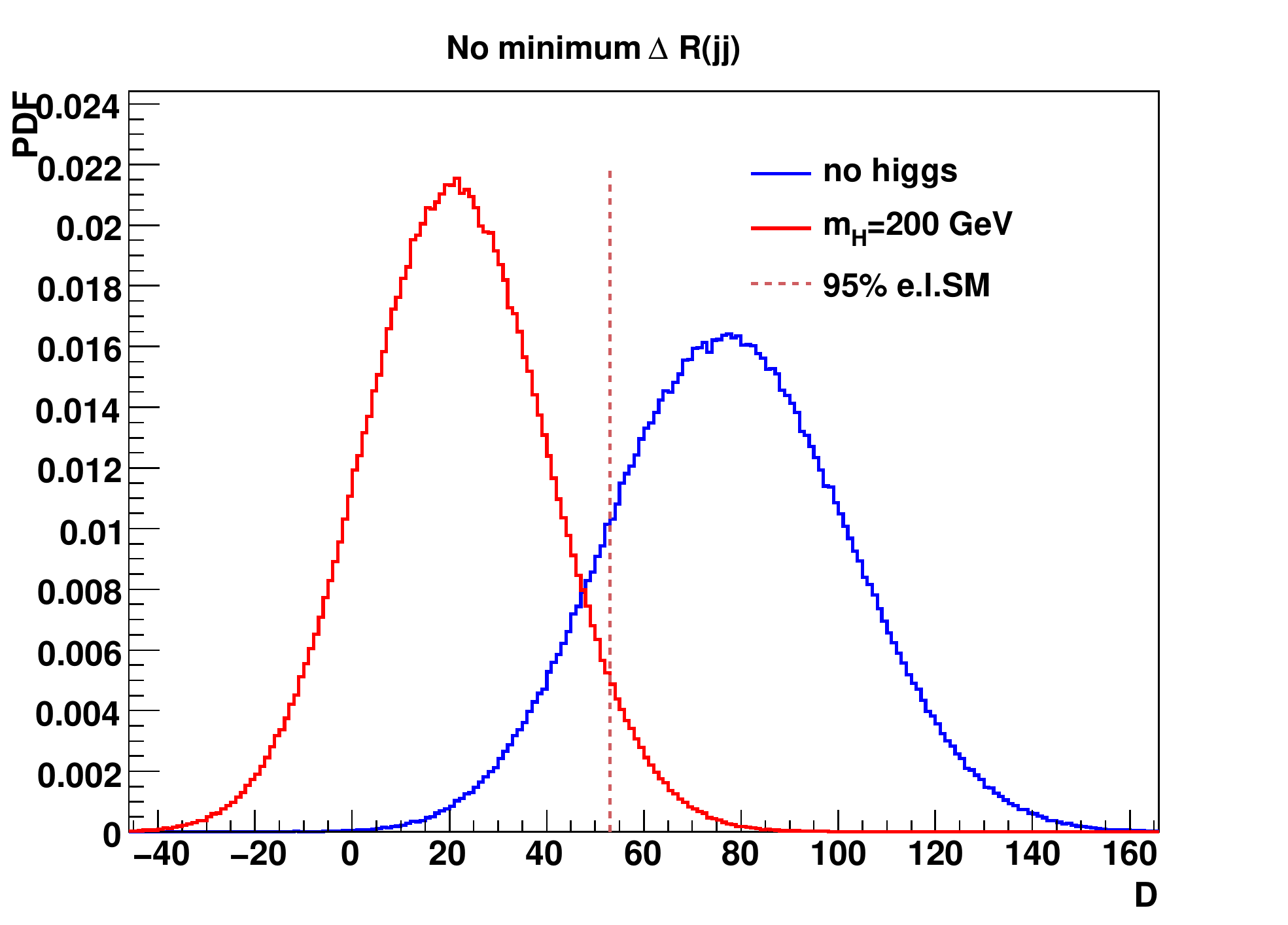}
\includegraphics*[width=0.32\textwidth,height=4.0cm]{mumu_pdf.pdf}
\includegraphics*[width=0.32\textwidth,height=4.0cm]{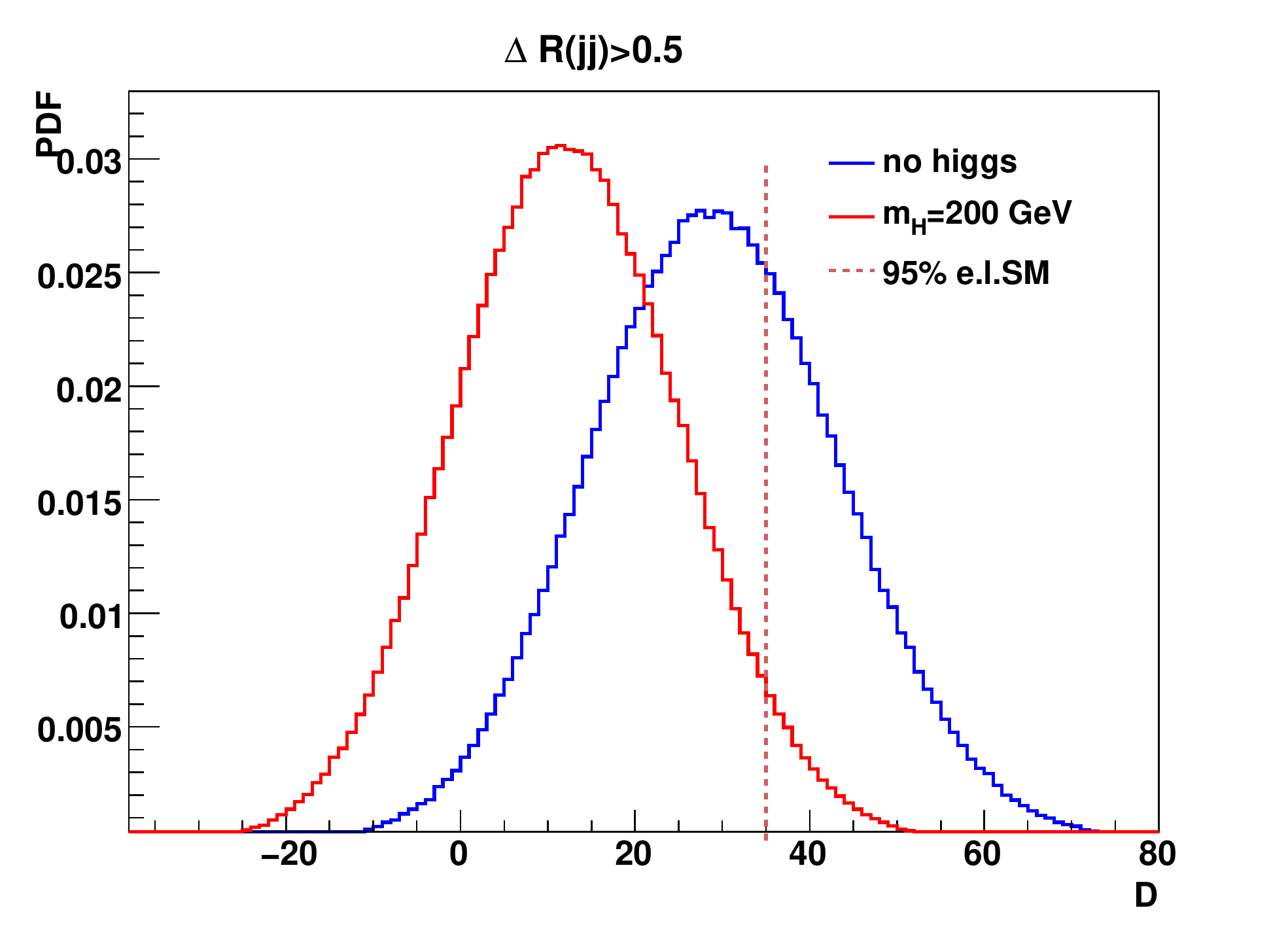}
\caption{Probability distribution of SM~(red) and \emph{No Higgs}~(blue)
  scenarios in the $4j\ell^+\ell^-$ channel, for different requirements on
  the minimum $\Delta R(jj)$ separation. From left to right: no $\Delta
  R(jj)$ requirement, $\Delta R(jj)>0.3$, $\Delta R(jj)>0.5$.}
\label{deltar}
\end{figure}

\subsection{Hadronic Decays of Vector Bosons and Jet Algorithms}
LVBS in the semi-leptonic channels contains a high-$p_T$ vector boson
decaying into two jets.  These two jets are very likely to be quite close in
$\eta$-$\phi$ space. Consequently, the choice of an appropriate jet algorithm
plays an important role for this kind of analysis.  We examine this problem
in Fig.~\ref{deltar}, where the probability distributions are shown for
different requirements on the minimum $\Delta R$ separation between jets. The
large loss in discriminatory power is evident as the distributions merge into
each other as the cone is enlarged.  This problem must be understood and
tackled in a hadronic environment, eventually testing alternative jet
algorithms other than the cone-based ones.

\subsection{Combined Channels}
\begin{figure}[t]
\centering
\includegraphics*[width=0.32\textwidth,height=4.0cm]{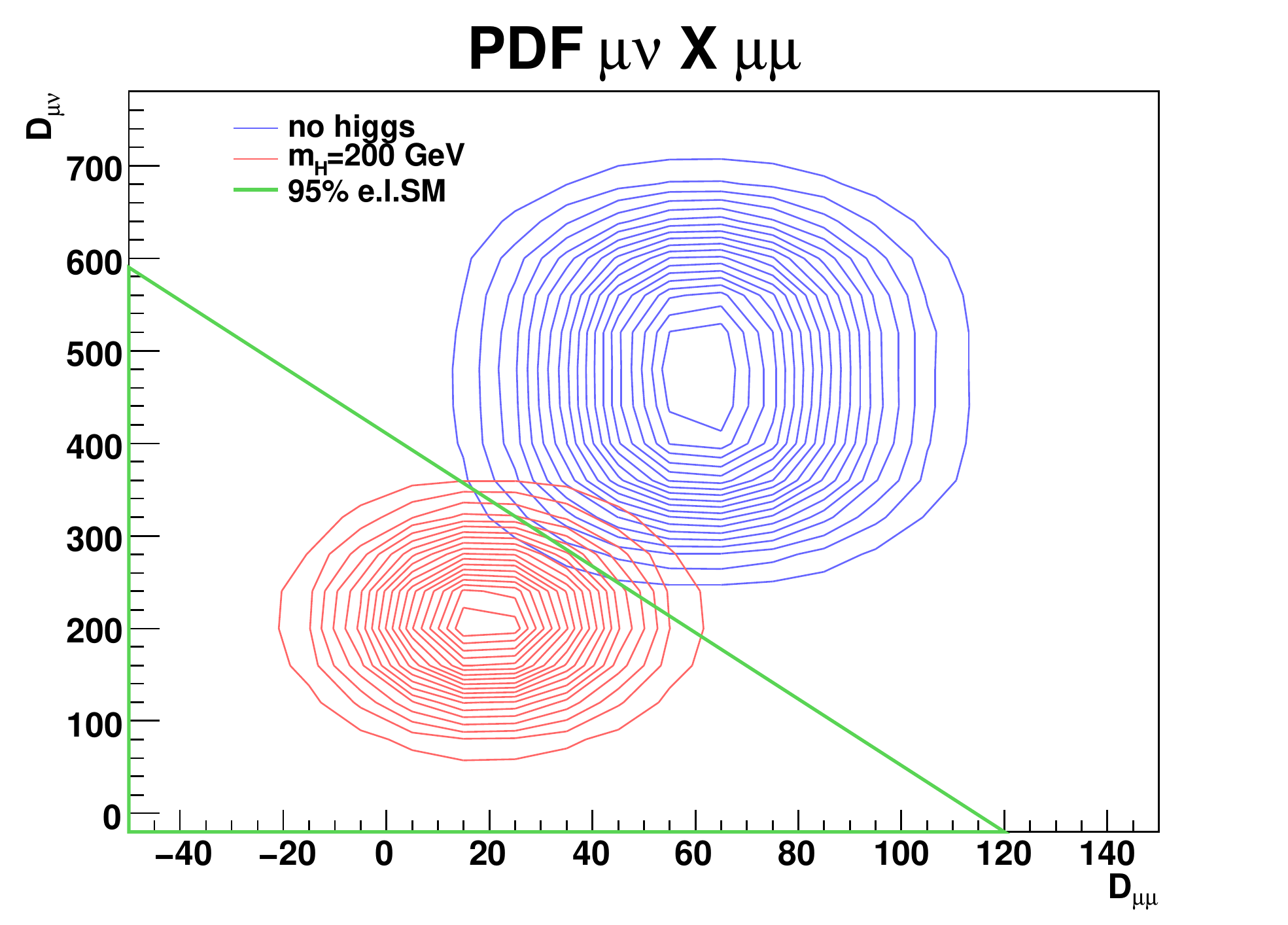}
\includegraphics*[width=0.32\textwidth,height=4.0cm]{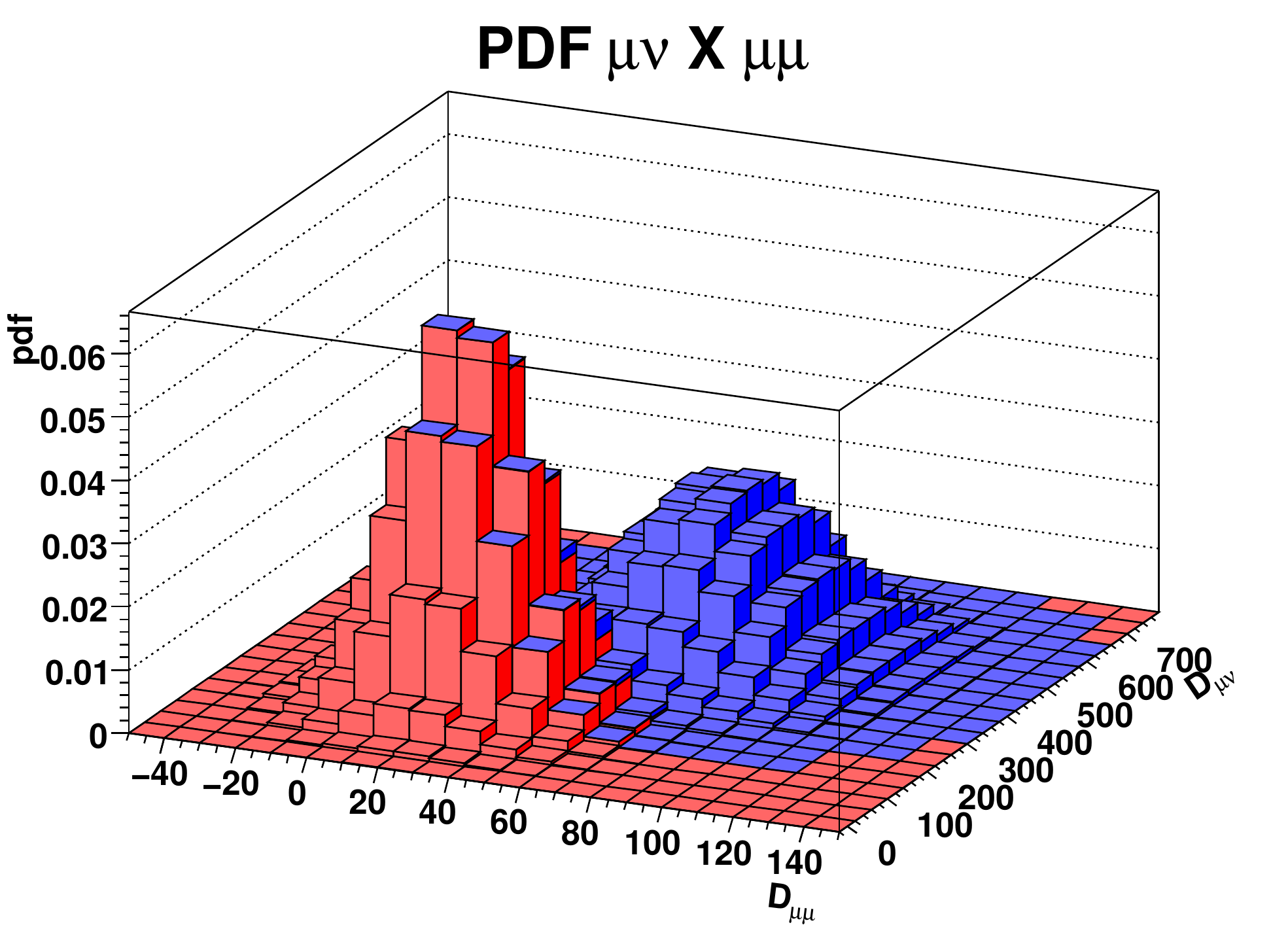}
\caption{Probability distribution of SM~(red) and \emph{No Higgs}~(blue)
  scenarios in terms of the discriminant of the semi-leptonic channels
  combined. The green line refers to the 95\% exclusion limit for SM.}
\label{comb}
\end{figure}

In order to have a better overall estimate of the discriminatory power of the
full analysis, it is useful to combine all channels. In fig.~\ref{comb}, the
procedure is illustrated for the combination of semi-leptonic channels.  In
order to define the surface that represents the 95\% exclusion limit of the
SM, we have used a criterion based on a likelihood-ratio test, separating the
points in which the ratio of the \emph{No Higgs} to the SM probability is
less than a certain value $a$.  The resulting PBSM@95\% CL are: $99.2\%$
combining the semi-leptonic channels and $99.9\%$ combining all three
channels.

\section{Conclusions}

We have estimated the power of the LHC in distinguishing the \emph{No Higgs}
scenario from the SM.  For this, we have performed a complete partonic
analysis generating full six-partons final states. We have developed a
strategy to suppress the main backgrounds using strong selection cuts, aiming
at optimizing the PBSM@95\% CL. With this approach we obtain a precise
probabilistic meaning of the discriminatory power, which has been
generalized to the combination of many channels. The final results show a
very good discriminatory power between the scenarios. The important issue of
resolving the hadronic decays of vector bosons must be better understood.

\end{document}